\documentstyle[11pt,psfig]{article}

\newcommand{\D}{\displaystyle}
\newcommand{\uu}{{\bf u}}
\newcommand{\xx}{{\bf x}}
\newcommand{\UU}{{\bf U }}

\newcommand{\kk}{{\bf k}}

\newcommand{\palp}{p_{\alpha_+}}
\newcommand{\qalp}{q_{\alpha_+}}
\newcommand{\pqalp}{{p_{\alpha_+}}^2 + {q_{\alpha_+}}^2}

\newcommand{\grad}{\nabla}

\newcommand{\ddiv}{\hbox{div}\,}
\newcommand{\xxpr}{{\bf x^{\prime}}}

\newcommand{\xxse}{{\bf x^{\prime\prime}}}

\newcommand{\wpr}{{\omega}^{\prime}}
\newcommand{\upr}{u^{\prime}}
\newcommand{\vpr}{v^{\prime}}

\newcommand{\uupr}{{\bf u^{\prime}}}

\newcommand{\dr}{\partial}

\newcommand{\EQ}{\begin{equation}}
\newcommand{\EN}{\end{equation}}
\newcommand{\EQA}{\begin{eqnarray}}
\newcommand{\ENA}{\end{eqnarray}}

\begin{document}

\title{ 
Fast Numerical simulations of 2D turbulence using \\
a dynamic model for Subgrid  Motions}

\author{J-P. Laval$^{1}$,  B. Dubrulle$^{2,3}$ and  S. V. Nazarenko$^{4}$}


\maketitle

\begin{enumerate}
\item[$^1$] CEA/DAPNIA/SAp, CE Saclay, 91191 Gif sur Yvette Cedex, France
\item[$^2$] NCAR, P.O. Box 3000, Boulder CO 80307-3000
\item[$^3$] CNRS, URA 285, Observatoire Midi-Pyr\'en\'ees, 14 avenue Belin, F-31400 Toulouse, France
\item[$^4$] Mathematics Institute University of Warwick COVENTRY CV4 7AL, UK
\end{enumerate}

\begin{abstract}

We present numerical simulation of 2D turbulent flow using a 
new model for the subgrid scales which are computed using
 a dynamic equation linking the subgrid scales with 
the resolved velocity. This equation is not postulated, but 
derived from the constitutive equations under the assumption that the 
non-linear interactions of subgrid scales
between themselves are equivalent to  a turbulent viscosity.
 This results in a linear
stochastic equation for the subgrid scales, which can be
numerically solved by a decomposition of
the subgrid scales into localized wave-packets.  These
wave-packets are transported by the resolved scale velocity
and have wavenumbers and amplitude which evolve according to the resolved
 strain
and the stochastic forcing. The performances of our model
are compared with Direct Numerical Simulations of decaying and forced 
turbulence. For a same resolution, numerical simulations using 
our model allow for a significant reduction of the computational time
(of the order of 100 in the case we consider), and allow the achievement 
of significantly larger Reynolds number than the direct method.
\end{abstract}


\vspace{0.1cm}


\section{Introduction}
\label{sec-intro}
The dynamics of high Reynolds number turbulent flow couples a large range of
 scales from the characteristic size of the domain to the 
dissipative scales.
 This range is usually too large to be fully resolved by Direct Numerical
 Simulation of the Navier-Stokes equations (DNS). This is the case for many
 applications like in aeronautics, geophysics or astrophysics where the
 typical Reynolds numbers are of the order of $10^6$ to $10^{12}$. The
 largest Reynolds numbers which can be achieved by DNS are of the
 order of $10^4$ to $10^5$~\cite{leonard95,werne99}. The difficulty 
of direct simulation of turbulent flows at high Reynolds number 
arises because of two scaling laws: first, the memory requirement 
to be able to deal with all the scale of motions growths like 
$Re^{9/4}$, where $Re$ is the Reynolds number; second, the time stepping
to be used to advance the equations has to be computed using the
smallest resolved scale. As a result, computational time usually growths
like $Re^3$. 
Despite of the 
growth of the computer power~\cite{joslin97}, the direct simulation
 of all realistic problems will not be conceivable before a long time.\

Part of these difficulties could be circumvented if one could find 
an efficient approximate scheme to model the small scales
of the turbulent flow, which usually monopolize 90 per cent of the 
computations and, for most applications, do not need to be known with
the same precision as the large scales. From an algorithmic point of view,
several methods have been proposed to try to describe 
the small turbulent scales with less degree of freedom than in a DNS like
for example, 
sparse Fourier transform \cite{grossmann96}, rarefied modes \cite{meneguzzi96},
wavelets \cite{schneider97}, self-adaptative grid mesh \cite{berger84,berger89}.
From a theoretical point of view, the effort has been put on the model
of the action of the small scales onto the larger scales of motion, or
onto the small scales themselves, to be able to deal with simpler and cheaper
systems to compute. In the extreme case of large eddy simulations (LES) or
random averaged numerical simulations (RANS), the effect of the 
small scales at large scales is directly modeled as a function of the
resolved or mean scales of motions, resulting in both memory and computational
gain. This is often done at the price of introduction of 
arbitrary parameters, which require calibration and which lower the predictive 
power of the simulation. Also, some models are not theoretically
satisfying, since they break for example the basic symmetries of 
the original equations \cite{oberlack97}.\

In this paper, we continue a study initiated in \cite{laval98} and develop a 
new dynamic model for the turbulent subgrid
quantities which is directly derived from the constitutive equations of
motions. This model  respects all the basic symmetries of the
original equations, as well as their conservation properties. 
The main assumption underlying the model is that the smaller turbulent scales
are much less energetic than the resolved scales, and so, that their main 
interactions are with the resolved flow. It then appears
reasonable to try to model these main interactions with as few approximations
as possible, while more freedom is allowed regarding their mutual
interactions. In the model we consider, the interactions between the resolved 
and subgrid scales are taken into account without any approximation, while
the mutual interactions between subgrid scales are 
replaced by a turbulent viscosity. In
2D geometry, where energy condensation at large scale guarantees 
that the dynamics of the small scales is mainly non-local in scales 
this turbulent viscosity is so small that it can be set to zero 
in some simple cases
\cite{laval99}. This allows a parameter free model of the subgrid
scales, which will be described in 
the first part of this paper. 
In situations where 
dynamic processes at small scales, like vortex stretching, are 
important to regulate any small-scale instability, or even in 
2D when "large" and "small" scales are not
well enough separated, the turbulent viscosity cannot be set to zero
\cite{laval00a}.
This option introduces a free 
parameter into the model, but, since it appears at a sub-dominant level,
we may expect that its choice is not so critical to the success of 
the model.

After parameterization of  the non-linear interactions 
between the subgrid scales, their resulting equation of motion becomes
linear. This essential feature has two advantages: first, it enables,
in certain simple flow geometries,
analytical solutions for the subgrid-scale dynamics as a function of the 
resolved quantities, thereby closing the equations of motions
at the resolved-scale level. This property was used to obtain analytical
solution for mean profiles in channel flows \cite{nazarenko99,nazarenko99b,dubrulle99b} or for
the Planetary Surface Layer \cite{dubrulle99a1,dubrulle99a2}. Second advantage of the
linear description is that it allows for more efficient algorithms of 
integrations, using Lagrangian methods where the time stepping is
done via a criterion based on the resolved scales. This allows numerical
computations with a larger time step than DNS, and thus, at a lower 
computational cost.\

An essential part of our model is the averaged Reynolds stresses describing the
feedback of the subgrid scales onto the resolved component. In \cite{laval98}, 
we considered the 
scale separation parameter to be much less than the nonlinearity of the small
scales and derived a feedback term in which the quadratic in the small-scale
amplitude terms gave the main contribution into the Reynolds stress. Then, the model
that consisted of two coupled ``fluids'' (the resolved and the subgrid one) was
used to solve several problems with (\cite{nazarenko98}) and without scale separation
(\cite{laval98}), the later being the typical problems such as the forced turbulence,
the vortex merger and a turbulence decay. It was noticed that in
the problems without a natural scale separation, such a model described very well
the small-scale dynamics whereas improvements in modeling of the large scales
(which is the main aim of the LES) was much more modest. The situation was clarified
in \cite{laval99} where direct numerical evaluations of the different contributions 
to the Reynolds stress were made. It was shown that in the problems without
the scale separation the dominant contribution to the Reynolds stress comes from 
the linear rather than quadratic in the small-scale amplitude term.

In the present contribution, we focus on advancing the numerical
approach introduced in \cite{laval98} by introducing a better model
for the turbulent Reynolds stress based on the results of the a priori tests
performed in \cite{laval99}. Also, a more efficient procedure is used in this
paper to generate the subgrid vorticity and velocity fields. 
Our goal here is to  test our model both as a numerical method
for an improved DNS, in the sense that we can compute the
whole range of scales at a lower computational cost, and as pseudo
LES method, in the sense that we compute only the larger scales, at an even
lower computational cost. For the sake of simplicity, we shall 
consider here only the two-dimensional case, where both 
the hypothesis of the turbulent model \cite{laval99} and its 
 consequences \cite{nazarenko98,laval98} have been studied in detail. 
The generalization to the 3D case is the subject of an ongoing research.

\section{The turbulence model}

\subsection{Scale decomposition}

We consider a two-dimensional incompressible inviscid
 flow obeying the equations:
\begin{equation}
\left \{
\begin{array}{lll}
\partial_t \omega + \hbox{div} (\uu  \omega) &=& \nu \Delta \omega, \\
\hbox{div} \, \uu &=& 0,
\end{array}
\right.
\label{euler2D}
\end{equation}
where $\uu$ is the velocity
and $\nu$ is the viscosity. In 2D geometry, the vorticity has 
only one non-zero component 
that we denote by $\omega$. 
The resolved-  and the subgrid-scale part of the
 velocity and the vorticity fields are defined via a
 filtering procedure:
\begin{eqnarray}
\UU(\xx,t) &=& \overline{ \uu(\xx,t)} = \int G(\xx-\xxpr) \uu(\xxpr,t) d\xxpr \\
\Omega(\xx,t) &=& \overline{ \omega(\xx,t)} = \int G(\xx-\xxpr) \omega(\xxpr,t) d\xxpr
\label{eq:deffilter}
\end{eqnarray}
Here, $G$ is a filter, such that 
 the resolved scales
 contain the main part of the total energy. In Section III.C, we shall 
propose a special shape of $G$ obeying this condition, and which
is well adapted to our numerical method.\

Each field is then
 decomposed as follows
\begin{eqnarray}
\uu(\xx,t)  &=& \UU(\xx,t) + \uupr(\xx,t), \\
\omega(\xx,t) &=& \Omega(\xx,t) + \wpr(\xx,t),
\end{eqnarray}
where the primes denote subgrid-scale quantities.
Inserting this decomposition into the Navier-Stokes
 equations and separating the resolved
 and the subgrid-scale parts, we get a set of coupled equations:
\begin{eqnarray}
\partial_t \Omega &+&\ddiv(\overline{\UU \Omega}) + 
\ddiv(\overline{\uupr \Omega}) \nonumber\\
&+&\ddiv(\overline{\UU \wpr}) + \ddiv(\overline{\uupr \wpr})=\nu \Delta\Omega,
\label{eq:decompls}
\end{eqnarray}
and
\begin{eqnarray}
\partial_t \omega' &+& \ddiv (\UU \Omega) - \ddiv(\overline{\UU \Omega})
 \nonumber\\
&+&\ddiv (\UU\wpr ) - \ddiv(\overline{\UU\wpr})\nonumber\\
&+&\ddiv (\upr \Omega) -\ddiv(\overline{\upr \Omega}) \nonumber\\
&+&\ddiv (\upr \wpr) -\ddiv(\overline{\upr \wpr}) = \nu\Delta\wpr.
\label{eq:decompss}
\end{eqnarray}
The second up to the sixth terms of the l.h.s. of (\ref{eq:decompss})
are the contributions due to non-local interactions between the
resolved and the subgrid scales.
The last two terms in the l.h.s. of (\ref{eq:decompss}) are the contributions
due to non-linear (local in scale space) interactions among the subgrid
 scales.
In the favorable case where most of the energy is at the resolved
 scales, these
last two terms can be expected to negligible with respect to, e.g., 
the second up to the fourth terms of the l.h.s. of (\ref{eq:decompss}).
Indeed, in a recent 
detailed numerical analysis of the system (\ref{eq:decompls}) and
(\ref{eq:decompss}), Laval et al \cite{laval99} 
showed that for a very steep filter 
(when the filter $G$ is a cut-off in the spectral space),
  the small-scale dynamics
(and thus the large-scale dynamics) is independent of the 
local interactions: when the later are neglected, the small-scale velocity
and vorticity field are not significantly 
changed, in both forced and decaying turbulence,
even after several eddy turn-over times \cite{laval99}. 
The analysis
has also shown that in 2D turbulence, the leading order contribution 
at both large and small-scale comes from the correlations involving
the large-scale velocity field, i.e. $\UU\Omega$ and $\UU\wpr$. The
next leading order contribution comes from the correlations between 
small-scale velocity and vorticity $\upr\wpr$, while the correlation 
between small-scale velocity and large-scale vorticity $\upr\Omega$
gives the lowest contribution. 

\subsection{A priori numerical estimates}
\begin{figure}[hhh]
\centerline{\psfig{file=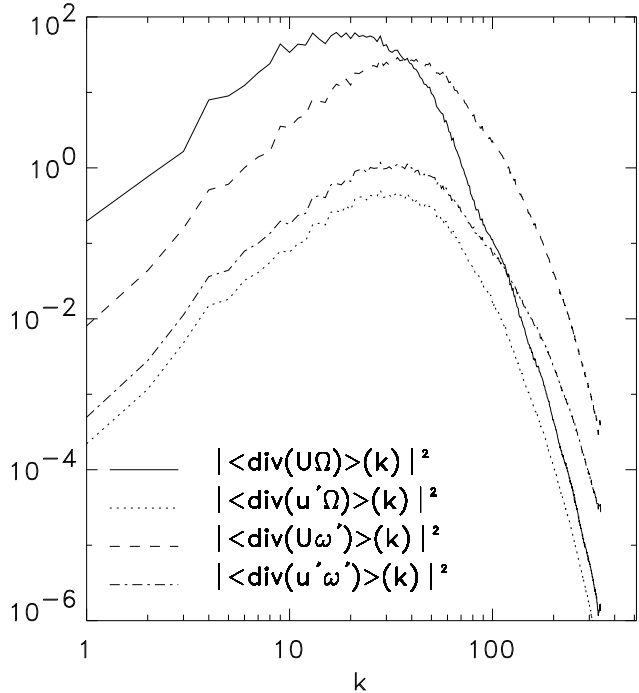,width=10cm}}
\caption[]{ Comparison of the square moduli of each component of the non-linear part of the large scale equation \ref{eq:decompls} in Fourier space. The filter used for the scale separation is a smooth filter in Fourier space defined by eq. (\ref{eq:filterfourier}). The energy spectra of both resolved scales and subfilter scales are shown fig. \ref{fig:testapriof2spct}}
\label{fig:testapriof2ls}
\end{figure}

\begin{figure}[hhh]
\centerline{\psfig{file=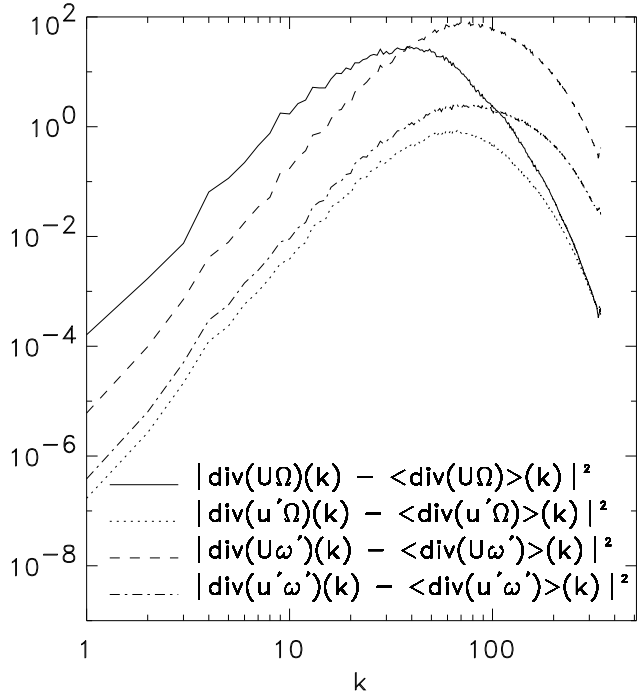,width=10cm}}
\caption[]{Same comparison as in fig. \ref{fig:testapriof2ls} 
but now for the non-linear terms in the subfilter scale equation.}
\label{fig:testapriof2ss}
\end{figure}

\begin{figure}[hhh]
\centerline{\psfig{file=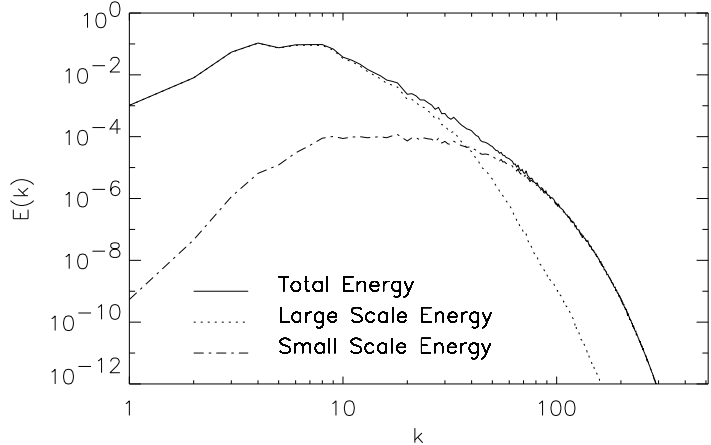,width=10cm}}
\caption[]{Energy spectra of the 
resolved and the  subfilter components
for the same run as in fig. \ref{fig:testapriof2ls} 
and \ref{fig:testapriof2ss}. 
The filter used for the scale separation is defined in 
eq. (\ref{eq:filterfourier}) with $dh=2 \pi /32$. 
The initial field is the result of a Direct Numerical Simulation of decaying turbulence on a $1024^2$ grid.}
\label{fig:testapriof2spct}
\end{figure}

%
In the present case, the filter $G$ is smoother, and subgrid scales 
include both large and small scales. It is thus important to conduct
additional numerical evaluation of the various 
terms of (\ref{eq:decompls}) and (\ref{eq:decompss}) using a smooth filter,
to check the validity of the non-locality assumption for the subgrid scales.
We performed an apriori test with the filter which will be used in our model. The choice of the filter will be discussed in section \ref{sec:filter}. The test was done using a vorticity field from a DNS on a $1024^2$ grid. The Figures \ref{fig:testapriof2ls} and \ref{fig:testapriof2ss} give the distribution of each non-linear terms involved in the resolved scales and the subfilter-scales equations. The result of the filtering process one the initial field is shown fig. \ref{fig:testapriof2spct}. Even with a smooth filter, the non-linear term involving only subgrid scales are still small compared to the higher order term.
%
%

\subsection{The model}
Keeping only the leading order 
contributions, the coupled equations
(\ref{eq:decompls}) and (\ref{eq:decompss})  become:
\begin{equation}
\partial_t \Omega +\ddiv \overline{\UU \Omega} +\ddiv \overline{\UU \wpr} 
 =  \nu \Delta\Omega, 
\label{eq:systls} 
\end{equation}

\begin{eqnarray}
\partial_t \omega' + \ddiv (\UU \wpr) & = & F
+\nu_t\Delta\wpr,\label{eq:systss} \nonumber \\
F(\xx,t) &= & -(\ddiv (\UU \Omega) -\ddiv \overline{\UU \Omega}) + 
\ddiv \overline{\UU \wpr}.
\label{eq:exprfss}
\end{eqnarray}
Here, $\nu_t$ is a turbulent viscosity, which will be introduced to
damp the small-scale noise arising in our numerical scheme and 
$F$ is a force which 
describes the subgrid-scale generation via the enstrophy cascade 
(energy cascade
in 3D).

\section{Numerical implementation}

\subsection{Numerical strategy and performances}

 The resolved-scale  equation has the form 
of an Euler equation with an additional forcing coming from the interaction 
with the subgrid-scale  motions. One is therefore led to standard strategies
to solve this equation, depending on the 
type of the flow
 (spectral methods for 
 flows in simple  periodic geometry, or
finite difference or finite elements 
for more complicated geometries). 
Here, we shall consider the periodic case, thereby using a spectral code
for solving the resolved-scale equation. This code is
described in \cite{brachet88}.\

The situation is markedly different at the subgrid level, where the basic 
equation is {\sl linear} in the subgrid motions, with an inhomogeneous
part provided by the subgrid scale generation via the enstrophy cascade.
This linearity suggests a solution strategy based on projection of
the subgrid scales onto appropriate modes. 
Since the subgrid scales are usually very inhomogeneous (e.g., vorticity
filaments),
it seems logical to use a decomposition of the
subgrid-scale velocity field into localized modes, thereby optimizing 
the memory requirements to store the subgrid-scale fields. A popular
local mode decomposition uses wavelets (see e.g. \cite{farge96}).
Wavelets are however sometimes difficult to implement, and are not
very handy to use in analytical computations. Here, we 
choose to use a Gabor decomposition, which provides a localized 
description while allowing theoretical manipulations similar to
that obtained with Fourier modes. 
The linearity of the subgrid motions also provides room
for a further  
computational time reduction via 
semi-Lagrangian methods of integration, 
using a time step related to the resolved scale.
The numerical expected performances of our model, integrated using
the semi-Lagrangian, Gabor method, at a given resolution and total 
integration time,
is given in Table \ref{tab:ordertime}, and compared with both standard spectral methods 
based on FFT, and a LES approach based on FFT.
%
%
\begin{table}
\begin{center}
\begin{tabular}{|l|c|}
\hline
\hspace{4mm} Model (resolution)       & integration time \\
\hline
DNS [FFT code] (N x N) &  $c_1 * N^2 log (N)$ \\
LES [FFT code] (M x M) &  $c_2 * M^2 log (M)$ \\
Our Model [FFT + Lagr.] &  $c_3 * M^2 log (M) + c_4 * N_p$ \\
\hline
\end{tabular}
\caption[]{Comparison of expected performance between DNS, Classical LES models (like APVM or HDNS) based on FFT on a (M x M) grid and our model on a (M x M) grid for fully resolved scales and $N_p$ modes for approximated subfilter scale equation (see section \ref{sec:lagrange}).}
\label{tab:ordertime}
\end{center}
\end{table}
Computational time for several particular examples is
given in table \ref{tab:cor} which shows an obvious gain
in speed for our method with respect to the DNS.
%
%
\begin{table}
\begin{center}
\begin{tabular}{|l|c|c|l|}
\hline
\hspace{4mm} Model       & \hspace{2mm} $\tau=15$ \hspace{2mm} & \hspace{2mm} $\tau=50$ \hspace{2mm} & Comp. time \\
\hline
DNS ($N=1024$)           &  ------    & ------    & $\sim$ 10 days \\ 
M0 ($M=64, N_p=512^2$)   &  0.99403   & 0.89794   & $\sim$ 1h30 \\    
M1 ($M=64, N_p=512^2$)   &  0.99450   & 0.82523   & $\sim$ 1h30 \\    
M2 ($M=64, N_p=128^2$)   &  0.99409   & 0.89735   & $\sim$ 8 mn \\    
APVM ($M=64^2$)          &  0.95882   & 0.77429   & $\sim$ 2 mn \\    
HDNS ($M=64^2$)          &  0.93893   & 0.53231   & $\sim$ 2 mn \\    
\hline
\end{tabular}
\caption[]{ Correlation between large-scale ($k<21$) vorticity from DNS
 and vorticity field from : our model with a maximum of $512^2$ particles
 (M0 and M1) and $128^2$ particles (M2), the APVM model and an hyper-viscous
 simulation both on the resolved-scale grid $N=64^2$. The same viscosity as in
 DNS was
 introduced in the resolved scale equation for the M1 simulation whereas
 no viscosity was
 introduced for the M0 and M2 simulations. The correlation coefficient
 are given at two different times (15 and 50 turnover times). The computations
 time with a ``Sun Ultrasparc 3000 workstation'' are given in the last column.}
\label{tab:cor}
\end{center}
\end{table}
If we now compare with a classical LES, our method is somewhat
slower because it is necessary to keep a sufficient number of
localized modes for accurate description of the Reynolds stresses.
However, a substantial gain in accuracy is achieved in comparison
with traditional LES because of the better description of the
nonlocal interaction of scales.

\subsection{Description of the algorithm}

Our algorithm of resolution of the turbulent model is based on five steps:
\begin{itemize}
\item[1] Compute the force $F$ (\ref{eq:exprfss}) using the resolved 
and subgrid fields at time $t$;
\item[2] Project this force onto a  set of Gabor modes;
\item[3] Advance the subgrid field to time $t+dt$ 
in the Gabor space using this projection and the Lagrangian 
advection algorithm;
\item[4] Compute the resolved Reynolds stresses at time $t+dt$ 
using a procedure similar to an inverse Gabor transform;
\item[5] Advance the resolved velocity field at time $t+dt$ using this
Reynolds stress.
\end{itemize}

Step 1 and 5 involve standard procedures linked with our resolved scale code
(in the present case a Fourier spectral code with an Adams-Bashforth 
time stepping where a full desaliasing was introduced by keeping only the first
 2/3 smaller wavenumbers in each direction, see
\cite{brachet88}). Step 2, 3 and 4 involve new original procedures,
based on interesting properties of the Gabor transform. For the sake of clarity
, we present here only main results, leaving detailed 
computations in appendices.

\subsection{The continuous Gabor transform}

The Gabor transform (GT) is defined as

\begin{eqnarray}
\hat{\uupr}(\xx,\kk,t) &=& \int f(\epsilon^*(\xx-\xxpr)) \, e^{i\kk \cdot (\xx-\xxpr)} \, \uupr(\xxpr)
 \, d\xxpr,
\label{eq:defgab}
\end{eqnarray}
where $f$ is a rapidly decreasing function at infinity. Note that 
$1/\epsilon^*$ has a meaning of the scale separation length. It has
to be chosen to lie in between of the integral scale and the Kolmogorov
scale, the later being far in the subgrid range in many applications.
For LES purposes, $1/\epsilon^* $ has to be close to (but not less than) 
the minimal resolved scale (grid scale $dh$ in our case).
Thus, to derive our model we need to assume the subfilter-scale
wavenumber $k$ to be greater than $\epsilon^*$ and use $\epsilon^*/k$
as a small parameter. Technically, however, is convinient to
keep $k$ fixed and perform expansions in small $\epsilon^*$.

In the special case where $f=\sqrt{G}$, where $G$ 
is the filter 
(see eq. (\ref{eq:deffilter})), we have the following simple reconstruction
formulae: 

\begin{equation}
\uupr(\xx,t) = \frac{1}{(2\pi)^2\;f(0)} \int \hat{\uupr}(\xx,\kk,t) d\kk
\label{eq:rebuiltx},
\end{equation}

\begin{equation}
 \overline{\UU \wpr}(\xx,t) = \UU \frac{1}{(2 \pi)^2} \int \Re
 \left[\hat{\wpr}(\xx,\kk,t) f(-\kk,t) \right] d\kk + O({\epsilon}^*),
\label{eq:defmoyp}
\end{equation}
where $\Re$ means the real part, and finally
\begin{equation}
\overline{u'_i u'_j}(\xx,t) = \frac{1}{(2 \pi)^2} \int \Re
\left[\hat{u'_i}(\xx,\kk,t) \hat{u'_j}(\xx,-\kk,t) \right] d\kk +
 O({\epsilon}^*).
\label{eq:defmoypp}
\end{equation}
One may also note the connection between the Gabor velocity and the Gabor
vorticity (similar to the Fourier quantities):
\begin{equation}
\hat{\uu}(\kk,t)  = -\frac{i}{k^2} \, \hat{\omega}(\kk,t) (e_z \times \kk)+
O({\epsilon}^*),
\label{eq:defvitug} 
\end{equation}
where $e_z$ is a unit vector in $z$-direction.\

Finally, one may show that the GT of
the subgrid-scale equation (\ref{eq:systss})
is
(see appendix \ref{sec:derivss} 
and \cite{nazarenko99,dubrulle99b} for more details)

\begin{equation}
 \left(\partial_t + \UU\cdot \grad_x -\grad_x\left(\UU\cdot \kk\right)\cdot
 \grad_k\right)
\hat{\wpr}(\xx,\kk,t) = \hat{F}(\xx,t)-\nu_t k^2 \wpr(\xx,\kk,t).
\label{eq:gaborss}
\end{equation}

\subsection{The discretized optimum Gabor transform}

The continuous Gabor transform is not suitable for numerical implementation.
Moreover, because it involves both position and wavenumbers, it
theoretically describes a given field via a much larger mode numbers than in
traditional spectral methods (typically $N^{2D}$ versus $N^D$ in D dimensions).
However, as discussed previously, inhomogeneous fields may be
represented with a high precision by a much smaller set of Gabor modes
than the theoretical number, a number even potentially smaller than $N^D$
(see e.g. \cite{farge99}). The algorithm we devised uses $N^D$ Gabor modes 
and, therefore, there is a room for further code optimization.

\subsubsection{The discretization}

The discretization of the subgrid field in the Gabor space is done
via a 
Particle In Cell (PIC) method. This method is
traditionally used in plasma physics, but was used previously 
in hydrodynamics by Nazarenko {\sl et al} \cite{nazarenko95} to study 
the interaction of sound wave-packets with turbulence. Details
about the method can also be found in \cite{laval98}.
In this method, the Gabor 
modes of the subgrid-scale vorticity field are 
replaced with discrete wave-packets (particles).
For example, the vorticity field 
$\hat{\wpr}(\xx,\kk,t)$ is discretized with $N_p$ wave-packets $\alpha$
 in the Gabor space:
\begin{equation}
\hat{\wpr}(\xx,\kk,t) = \sum_{\alpha=1}^{N_p} \hat{\sigma}_{\alpha}(t)
\; S_{\xx}(\xx-\xx_{\alpha}(t)) \; S_{\kk}(\kk-\kk_{\alpha}(t)).
\label{eq:rebuiltxk}
\end{equation}
In (\ref{eq:rebuiltxk}), $\alpha$ labels the wave-packet. Note that our
representation allows several wave-packets with different wavenumbers
to be located at the same position $\xx$, like in the Gabor transform. 
Here $S_\xx$ and $S_\kk$ are some interpolating functions (particle ``size'').
 Since the computation of non-linear terms in the resolved-scale equation 
involves a  second order spatial derivative,
 we are led to choose a linear interpolation
 in the $\xx$ direction. In the $\kk$ direction,
a zero particle ``thickness'' turns out to be sufficient for our purpose.
We thus adopt the following representation,
\begin{eqnarray}
S_\xx(\xx) &=& S(x) S(y),
\label{eq:picfuncx}\\
S_\kk(\kk) &=& \delta(p) \delta(q), 
\label{eq:picfunck}
\end{eqnarray}
where $\xx=(x,y)$, $\kk=(p,q)$, $\delta$ is the Dirac function and
the function $S(\eta)$ is defined by:
\begin{eqnarray}
\D S(\eta) & = & \left \{ 
        \begin{array}{ll}
          \D (dh  - |\eta|)/dh & {\rm if} \; |\eta|<dh,\\
          0 & {\rm otherwise}, \\
        \end{array} \right. 
\label{eq:defsx}\\
\end{eqnarray}
Here, $dh$ is a length scale governing the accuracy of the discretization.
Its choice will be discussed later on.

The PIC algorithm of reconstruction of the continuous fields 
from the discretized coefficients provides a natural filter $G$ and 
the function $f$ to be used in the Gabor transform
 through (see appendix \ref{sec:filter}):

\begin{equation}
G(\xx) =  f^2(\xx) = f^2(0) \times S^2_\xx(\xx),
\end{equation}

with 

\begin{equation}
f(0) = {1 \over \sqrt{\int S^2_\xx(\xxpr) d\xxpr}}.
\end{equation}
In the {\kk} space, the filtering operation is achieved
by simply multiplication of the Fourier coefficients by
the Fourier transform of $G$, which is
\begin{equation}
\hat G(p,q) = \frac{36}{(p\;dh)^2(q\;dh)^2}\left\{ \left(1-\frac{sin(p\;dh)}{p\;dh}\right)
\left(1-\frac{sin(q\;dh)}{q\;dh}\right) \right\}
\label{eq:filterfourier}
\end{equation}

\subsubsection{Reconstruction of the subgrid scale correlations}

With our discretization, the formulae of reconstruction of the 
subgrid correlations (\ref{eq:defmoyp})
can be obtained using (\ref{eq:rebuiltxk}).
 Taking into account the condition 
that $\wpr$ is real, we have (see appendix \ref{sec:numreynolds} for details):
\begin{equation}
\overline{\UU \wpr}(\xx,t) \simeq  2 \UU  \sum_{\alpha_+=1}^{N_p} \Re
 \left[\hat{\sigma_{\alpha_+}} f^*(\kk_{\alpha_+}) \right]
 S_\xx(\xx-\xx_{\alpha_+}), 
\label{eq:rebulswss}
\end{equation}
where the sum is only over the wave-packets with positive wavenumber
$p_\alpha$\footnote{.
 Since the  vorticity $\wpr(\xx,t)$ is real, each wave-packet 
at position $\xx_\alpha$ and with wavenumber $\kk_\alpha$ will
have a partner
 wave-packet with same amplitude, at the same location and with an 
opposite wavenumber $-\kk_\alpha$.}.\
Similarly, one can use  (\ref{eq:defvitug}) and  (\ref{eq:rebuiltxk})
to re-write  (\ref{eq:defmoypp}) as (\ref{eq:rebussvss}) (see appendix \ref{sec:numreynolds})
\begin{eqnarray}
\nonumber \overline{\upr \vpr}(\xx,t) = 2 \sum_{\alpha_+=1}^{N_p} \frac{-
\qalp \palp }{(\pqalp)^2} |\hat{\sigma}_{\alpha_+}|^2
S^2_\xx(\xx-\xx_{\alpha_+})\\
\overline{{\upr}^2}(\xx,t) = 2 \sum_{\alpha_+=1}^{N_p} \frac{+
\qalp^2 }{(\pqalp)^2} |\hat{\sigma}_{\alpha_+}|^2  S^2_\xx(\xx-\xx_{\alpha_+})
\label{eq:rebussvss} \\
\nonumber \overline{{\vpr}^2}(\xx,t) = 2 \sum_{\alpha_+=1}^{N_p} \frac{+
\palp^2 }{(\pqalp)^2} |\hat{\sigma}_{\alpha_+}|^2  S^2_\xx(\xx-\xx_{\alpha_+}) 
\end{eqnarray}

\subsubsection{Optimum discrete Gabor transform}

For a given field $\wpr(\xx,t)$, 
associated with a grid of size $2\pi/N$,
our "optimum" discrete Gabor transform
only retains one wavenumber per position (i.e. $N^2$ Gabor modes, 
for a $N^2$ initial discrete field). This wavenumber, and the amplitude
of the corresponding Gabor mode are chosen as follows:
we use the exact relation 
between the field and its $N^2$ wave-packet components  
\begin{equation}
\wpr(\xx,t) = \frac{2}{f(0)} \sum_{\alpha_+=1}^
{N_p} \Re
\left[ \sigma_{\alpha_+}(t) \right] S_\xx(\xx-\xx_\alpha).
\label{eq:forcess}
\end{equation}
An easy way to satisfy (\ref{eq:forcess}) exactly 
is to create one wave-packet at each grid point, so that $N_p=N^2$.
 Applying equation  (\ref{eq:forcess}) at each grid point $\xx_i$
 we then obtain:

\begin{equation}
\wpr(\xx_i,t) =  \frac{2}{f(0)}  \Re \left[ \sigma_{\alpha_i}(t) \right]. \\
\end{equation}

This equation fixes the real part and the physical coordinates of the 
wave-packet. To find its wavenumber, we use the velocity at the grid 
point $\upr(\xx_i)$ to find two conditions: 
\begin{eqnarray}
\upr(\xx_i,t) & = &  \frac{2}{f(0)}
\frac{-q_{\alpha_i}}{(p_{\alpha_i}^2 + q_{\alpha_i}^2)} \Im
\left[ \sigma_{\alpha_i}(t) \right], \\
\vpr(\xx_i,t) & = &  \frac{2}{f(0)} \frac{+p_{\alpha_i}}{(p_{\alpha_i}^2 +
 q_{\alpha_i}^2)} \Im
\left[ \sigma_{\alpha_i}(t) \right] .
\end{eqnarray}
These conditions link the two components of the wavenumber with the 
imaginary part of the amplitude of the wave-packet, which is still 
a free parameter at this stage. We then select the phase of
the GT by requiring
that the imaginary part of the wave packet equals its real part. Other choices
could have been made, but this one turns out to simplify the computations. 
The characteristics of each wave-packets $\alpha_i$ 
can finally be summarized as:

\begin{equation}
\left\{
\begin{array}{lll}
\D \Re \left[ \sigma_{\alpha_i} \right] & = & f(0)
\, \wpr_f(\xx_i) / 2 ,\\
\D \Im \left[ \sigma_{\alpha_i} \right] & = & \Re \left
[ \sigma_{\alpha_i} \right], \\
\D q_{\alpha_i}/p_{\alpha_i} & = &  - \upr_f(\xx_i) / \vpr_f(\xx_i), \\
\D p_i & = & \frac{ \wpr_f(\xx_i) \left( q_{\alpha_i}/p_{\alpha_i}
\right)^2}{\vpr_f(\xx_i) (1 + \left(q_{\alpha_i}/p_{\alpha_i}
\right)^2)} ,\\
\D \xx_{\alpha_i} & = & \xx_i.
\end{array} \right. \\
\label{eq:comppart}
\end{equation}
This procedure creates $N^2$ wave-packets, from any vorticity field 
on a grid of size $2\pi/N$ and can be used to initialize the subgrid
scale vorticity field from any given initial condition or to GT the force $F$
by transforming it into an equivalent vorticity field 
$\wpr_f(\xx,t)=F(\xx,t)/dt$.

\subsection{The Lagrangian scheme}
\label{sec:lagrange}

A Lagrangian interpretation of the subgrid scale
equation (\ref{eq:gaborss}) shows that its time integration
is equivalent to 
the evolution
 of each wave-packet in the (\xx,\kk) space. Each wave-packet carries 
 a complex amount of Gabor vorticity $(\sigma_\alpha)$ and is advected at the
resolved-scale velocity $\UU(\xx)$, while its wavenumber  and amplitude 
evolve according 
to the local resolved strain. For the trajectory of the wave-packet and
its amplitude we have
\begin{eqnarray}
\dot \xx_{\alpha}& = & \UU(\xx_{\alpha}(t)), 
\label{eq:eqdifpx} \\
\dot \kk_{\alpha}& = & - \grad_x\left(\kk_{\alpha}(t) \cdot
\UU(\xx_{\alpha}(t))\right), 
\label{eq:eqdifpk}\\
\dot \sigma_{\alpha}& = & \hat{F}(\xx_\alpha,\kk_{\alpha},t)-\nu_t k_\alpha^2
\sigma_\alpha.
\label{eq:eqdifps}
\end{eqnarray}   
By these equations, the wave-packets evolve continuously in the physical
space in between the grid points associated with the finite, resolved-scale
resolution. To find the resolved-scale quantities at the position 
of each particle (i.e. possibly in between grid mesh points), an
interpolation procedure is used via the functions 
$S_\xx$ and $S_\kk$ associated with the PIC method.

\subsection{Initial conditions}

All the simulations used the same initial random vorticity field with all the energy concentrated at very large scales ( the initial energy spectrum is given by $E(k) = k e^{-(k-k_o)^2}$ with $k_o = {\sl 1}$).
In practise, 
the noise associated with the PIC method is a function of the number
of wave-packets used for computation. If the initial condition is
such that the initial grid on which the vorticity is defined 
is too coarse ($M$ too small), one can use a two step procedure
to reach a reasonable number of wave-packets:
in a first stage of the simulation, 
we create at each time step $M^2$ wave-packets
with the vorticity $\wpr_f(\xx,t)=F(\xx,t)/dt$ 
created by the forcing $F$. During this stage, the  
wave-packets which were created at earlier times 
 are moved into the phase space using the ray 
equations (\ref{eq:eqdifpx},\ref{eq:eqdifpk}), 
but their amplitude is kept constant. This 
procedure is used until the total number of wave-packet reaches a desired
number. From this point on, the wave-packet creation is shut down, and
the wave-packets are evolved according to  
(\ref{eq:eqdifpx},\ref{eq:eqdifpk},\ref{eq:eqdifps}).\
   
\subsection{Noise reduction and effective turbulent viscosity}

For inviscid simulations (when the turbulent
viscosity is taken equal to $0$), the numerical
procedure  develops  a noise at very 
subgrid scales, which is a function both of the integration time and
the total number of particles. We have found that an 
efficient way to reduce this subgrid-scale noise could be achieved 
by periodically recreating a new set of $N_p$ wave-packets via first 
a rebuilding of the vorticity field in the physical space, on a 
grid of size $\sqrt{N_p}$ using 
(\ref{eq:rebuiltx}), 
followed by a re-creation of wave-packets using 
 (\ref{eq:comppart}). 
This procedure keeps the correlations 
associated with the subgrid-scale field unchanged, and therefore
does not affect directly the resolved scale. 
In effect, this
 procedure acts as a 
filter for the subgrid scales which are 
smaller than the size of the reconstruction
 grid, and thus, can be seen as an effective turbulent viscosity,
which cannot be estimated a priori but which adjusts itself to the noise
level.

\subsection{Test of the accuracy of the method}

\begin{figure}[hhh]
\centerline{\psfig{file=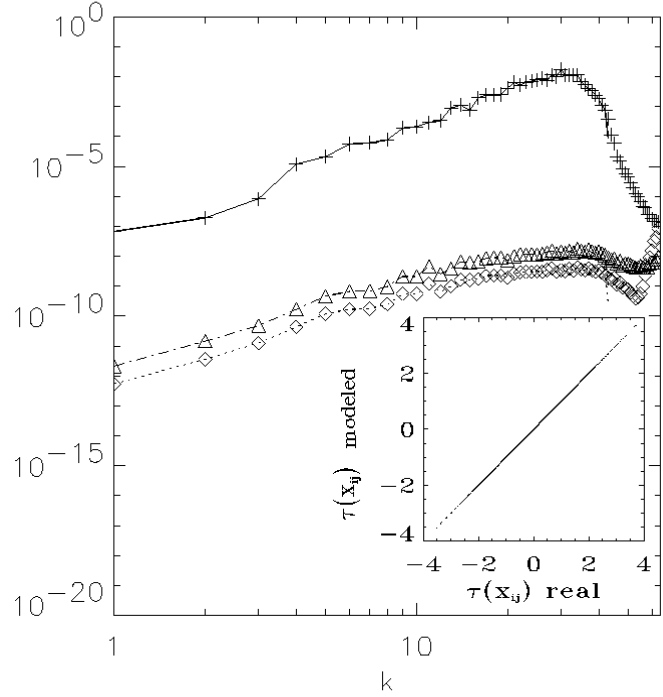,width=10cm}}
\caption[]{ Moduli of the 3 non-linear terms in
 the resolved-scale equation (\ref{eq:decompls}) computed directly
by the spectral method (lines) compared with the same fields rebuilt
by the PIC method (symbols).
Here, $+$ and ----- are used for the field
$\vert \ddiv \overline{(\UU \wpr)}(k) \vert^2$, 
$\triangle$ and $- \cdot -$ for
 $\vert \ddiv \overline{(\uupr \wpr)}(k) \vert^2$ 
and $\Diamond$ and $- - -$ for $\vert \ddiv \overline{(\uupr \Omega)}(k) \vert^2$). Corresponding comparison of the true and the PIC-modeled  
Reynolds Stresses
 $\tau = \ddiv(\overline{\uu\omega}) - \ddiv(\overline{\UU\Omega})$ 
at each grid point $X_{ij}$ is shown in insert.}
\label{fig:testapriok}
\end{figure}

To test the accuracy of our numerical method, we have performed 
{\sl a priori} tests using data from a DNS at high resolution.
 The total vorticity field  was
 divided into a subgrid-and a resolved-scale field using the same filter than
 in our model. The subgrid-scale field is further  discretized into
 wave-packets according to eq. (\ref{eq:comppart}).
A comparison was then made between the various components of the 
stresses 
 in the DNS, and in our discretization scheme.
 The result of
 is shown fig. \ref{fig:testapriok}, which shows the square
 modulus of each of the three Reynolds stress components with respect to
 their wavenumbers. The accuracy of the discretized reconstruction 
of each  
terms is remarkable  for the large scales. At very
small scales ($k>42$), the discretization 
tends to produce a noise in the components of the Reynolds stress 
involving the subgrid-scale velocity. In our case, we do not consider these 
terms so that we do not have to bother about this noise. 
In situations where one, or two of these troublesome terms are considered,
the noise can be removed via a filtering at small scale.
In spectral simulations, this truncature is provided naturally 
via the procedure to remove aliasing, which filters out modes larger than 
 $2 k_{max} / 3$ where $k_{max}$ is the maximum wavenumber of the
 resolved-scale field. 
%
%
%
%
 The inset of figure \ref{fig:testapriok} shows the comparison 
between the total real Reynolds stresses versus the total 
modeled Reynolds stress after desaliasing.
The agreement is nearly perfect. 


\clearpage
\section{Numerical results}

Our method can be used to perform two kind of simulations, depending
on whether one is interested in the small-scale behavior or not.
In the first case, one needs to model the subgrid scales with a large
number of wave-packets, so as to be able to reconstruct the small-scale
field with a good accuracy. Typically, one needs about $N_p=N^2$ 
wave-packets to be able to reconstruct faithfully  details at scale
$2\pi/N$ and  to produce 
a result which may be compared with a 
Direct Numerical Simulation at resolution $N^2$. 
In the case when only large scales matter, one needs to keep
the smallest number of wave-packet necessary to compute accurately 
the Reynolds stresses at the wavenumber cut-off. Since the resolved-
scale field at the cut-off $k_c = M/2$ produces (by non-linear beating)
information up to scale $2\pi/M$, we used a minimum of
$N_p=M^2$ wave-packets in our "Large Eddy Simulations". Finally,
note that our method allows nearly inviscid computations, since
it does not require the existence of a viscosity at large scale
and since it uses a minimal "effective viscosity"  which starts 
acting only at scales $2\pi/M$. Very large Reynolds number can
then be achieved via an adequate number of wave-packets.\

We present results illustrating these points in two classical 
situations,  - decaying and forced turbulence. In each case, the 
performance of our model are discussed and compared with 
results from the DNS and other popular subgrid-scale 
parameterizations used in 2D turbulence.

\subsection{Decaying turbulence}

For the decaying case, the initial condition was chosen 
so that the energy 
is concentrated at very large
 scales. 
 The reference DNS was performed at a resolution $N^2=1024^2$ with a viscosity 
 $\nu=1.8 \; 10^{-4}$, 
leading to a Reynolds number $Re\sim 10^4$. The simulation was 
stopped after approximatively 50 turnover times, by which time 
the initial condition has evolved into a robust dipole structure.
The separation 
  between resolved and subgrid scale used for our model
is taken at 
$k_c = 21$, corresponding to a computation over a grid 
$64^2$.\

Let us first compare the DNS to our model using a large number of wave-packets
($N_p = 512^2$). We ran two different simulations: one in which
the viscosity at resolved scale was set to $\nu=1.8\; 10^{-4}$, like in
the DNS ("viscous" simulation); another one in which $\nu$ was set to zero at 
resolved scale ("inviscid" simulation). The total simulation time 
of each simulation is about 1h30 on a Sun workstation, 
roughly 150 times less than the DNS.

\begin{figure}[hhh]
\centerline{\psfig{file=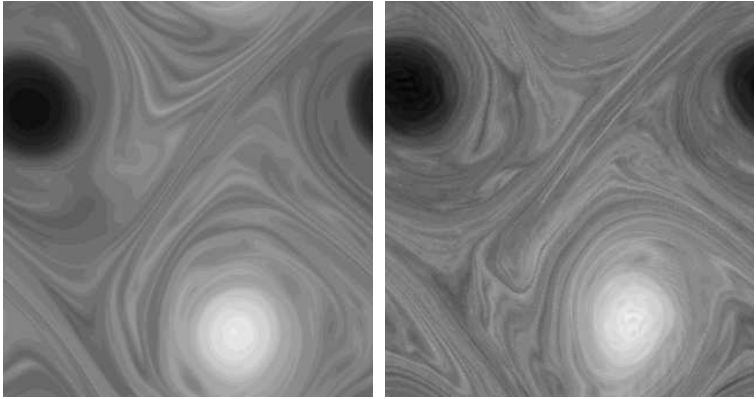,width=10cm}}
\caption[]{Total vorticity field of decaying turbulence
after 50 turnover times as computed by DNS on a $1024^2$ grid with viscosity
 $\nu=1.8 \; 10^{-4}$ (left) compared with the same field
computed by the model M0 described in table 2 (right).}
\label{fig:dtvortts}
\end{figure}

\begin{figure}[hhh]
\centerline{\psfig{file=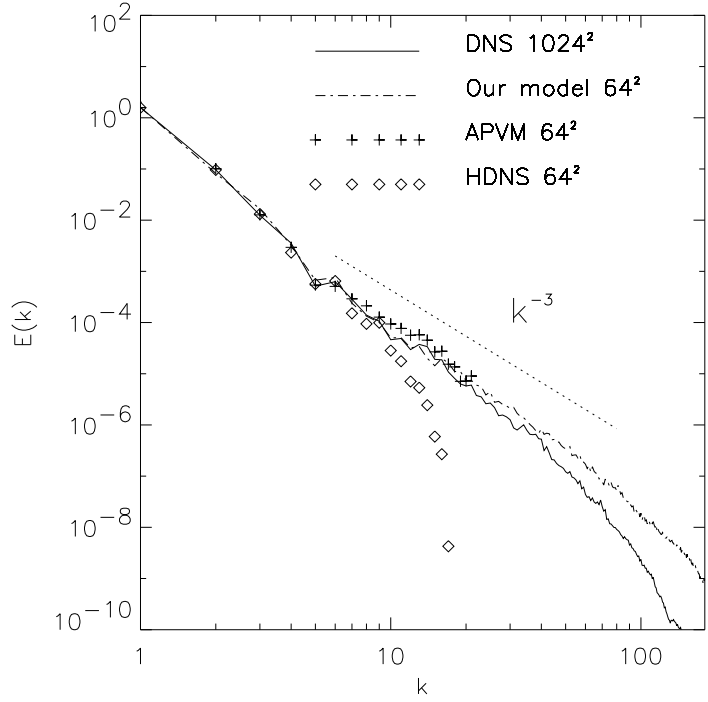,width=10cm}}
\caption[]{The energy spectra obtained
by the same simulation as in fig. 2 compared with 
the spectra obtained by some other methods including our 
model with parameters  M0 from the table 2.}
\label{fig:dtspect}
\end{figure}

 The total vorticity field after
50 turn over times,
reconstructed by adding the contribution from the resolved and subgrid scales,
is shown in Fig. \ref{fig:dtvortts} and compared with the vorticity field of
the DNS. 
A global comparison for small and large scales can also be done 
via energy spectra. This is done in Fig. \ref{fig:dtspect}. 
%
%

\begin{figure}[hhh]
\centerline{\psfig{file=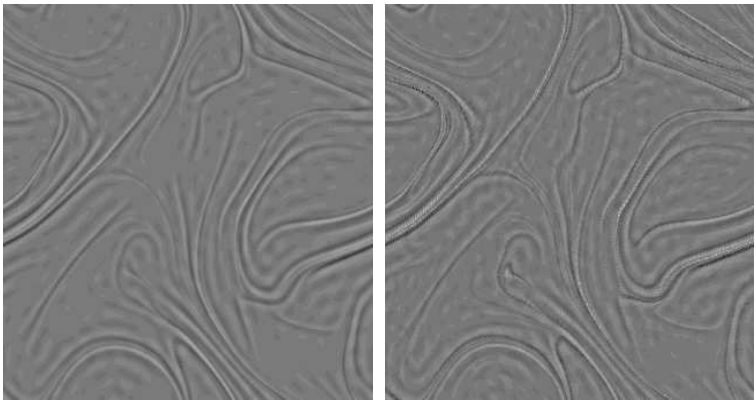,width=10cm}}
\caption[]{Small-Scale vorticity field of decaying turbulence
after 15 turnovers. Result from a DNS on a $1024^2$ grid with
 $\nu=1.8 \; 10^{-4}$ (left) and from our model (M0).}
\label{fig:dtvortss}
\end{figure}

%
%
Clearly, the two pictures display good overall similarities at large 
scales, and marked differences at smaller scales.
 The largest structures are very similar and they are well
 localized even after 50 turnover times. The spectra in the two simulations
overlap. The viscous simulation
gives large-scale structures which are in closer agreement 
with the (viscous) DNS (see Table 2). Together, these results confirm that 
at large scale, the dynamically important coupling term between 
resolved and subgrid scale is the term $div{(\overline U\omega')}$,
in agreement with the dynamical analysis of Laval et al
\cite{laval99}, performed using 
a cut-off filter.
At smaller scales, our model 
seems to produce thinner filaments and smaller structures. This effect
is clearly visible on the spectra of the two simulations: in the DNS, 
the $k^{-3}$ inertial law starts to level off towards $k=60$ (due to
viscous effects), while in our model, the power law extends over a wider 
range of scales, up to approximately $k=100$. To check whether this difference
 comes from our subgrid scale scheme, we performed a 
direct comparison of the 
smallest scales of the simulation $k>21$. Fig. \ref{fig:dtvortss} shows such a comparison
at some earlier time (after about 15 turnover times), when the small
scales have not yet been washed out by viscosity.
%
%
%
 At this time, 
differences are negligible, proving that our model also captures the 
dominant coupling mechanism at subgrid scales. This finding is also
in agreement with the dynamical analysis of Laval 
et al \cite{laval99}. The further differences arising over longer time scales
can therefore be due to two effects: one is the 
error accumulation due to  sub-dominant
 neglected terms in our model (like the terms 
involving coupling of the resolved vorticity $\Omega$ with the 
subgrid velocity $\uu'$); the second is viscous effects.  We believe that the
first possibility is ruled out by the dynamical analysis of Laval et al
\cite{laval99}, in which numerical simulations of our model equations
were performed using the same viscosity than in the direct numerical simulations
\footnote{These simulations were not fast: they were based on spectral methods
and were even more slowly than the direct method.}. In that case, the 
differences at small scales appear to be negligible. We therefore interpret
the differences between the two models as a viscous effect, and, actually,
as an indication of the lower effective viscosity in our model. This 
would explain both the later bending of the spectra, and the finer 
structure of the filaments.\

\begin{figure}[hhh]
\centerline{\psfig{file=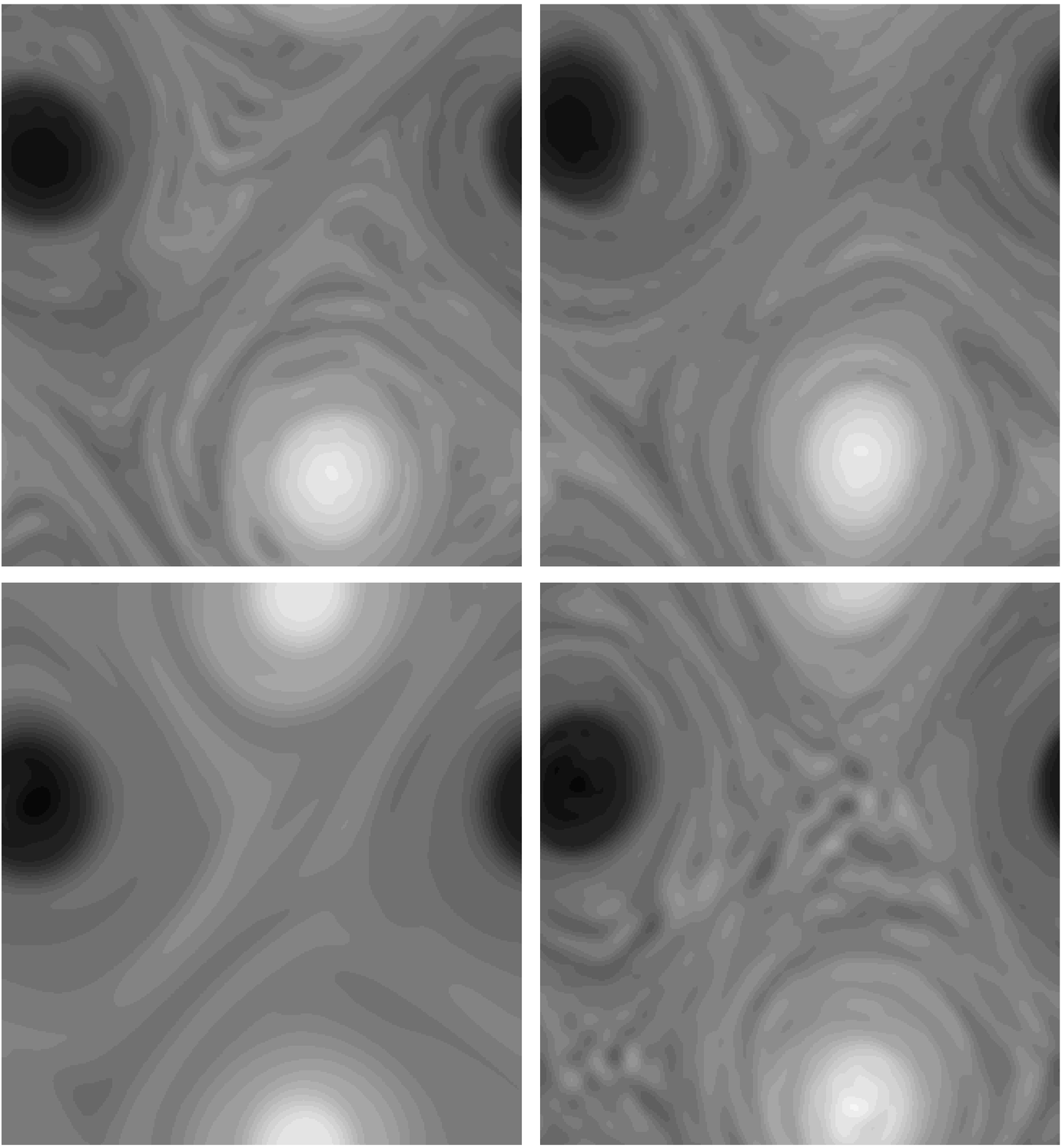,width=10cm}}
\caption[]{Large-scale vorticity field ($k \le 21$) after 50 turnover
 times in decaying turbulence. {\bf upper left}: DNS on a $1024^2$ grid and
$\nu=1.8 \; 10^{-4}$ , {\bf upper right}: our model (M2 in table 2)
 with a separation
 scale at $k=21$ and 16384 modes in subgrid scales, {\bf lower left}: 
simulation
 with hyper-viscosity (using $\nu_p k^p \omega$ as the 
dissipation term with p=8) on a $64^2$ grid and {\bf lower right}: simulation with the APVM model (\cite{sadourny85})
on a $64^2$ grid.}
\label{fig:dtvortls}
\end{figure}

The second series of comparison were performed using the minimum number
of wave-packet, equal in the present case to $N_p = 128^2$. In such a case,
the number of wave-packets does not allow  accurate reconstruction 
for the scales less than $2\pi/128$. The simulation is
however much faster, and takes only 8 minutes of computational time.
The total large-scale vorticity field after 50 turn over times is
compared with the corresponding large-scale vorticity field of 
the DNS in Fig. \ref{fig:dtvortls}. 
%
%
%
%
The agreement is still
very good and to quantify this agreement,
we computed the correlation coefficient between the two simulations,
defined by:  
\begin{equation}
C_{1,2} = \frac{\sum_{1<i,j<64} \omega_1(i,j) \omega_2(i,j)}
{\sqrt{\sum_{1<i,j<64} \omega_1(i,j)^2 \sum_{1<i,j<64} \omega_1(i,j)^2}}
\end{equation}

where $\omega_1(i,j)$ and $\omega_2(i,j)$ are the large-scale vorticity
field. This correlation coefficient was computed at two different 
times, corresponding to 15 and 50 turnover time, 
 and reported in table \ref{tab:cor}. At earlier time, our three models 
(viscous or not) are all characterized by a very good correlation
coefficient (about 99 percent). At later time, a slight difference 
appear. In fact, the best correlations are achieved by the viscous model,
and the model with the least number of subgrid-scale modes. This is not 
surprising, since our noise removing procedure produces an effective viscosity
which is larger as the number of subgrid-scale modes decreases. The 
high resolution model M1 is then the less viscous model of all three,
and therefore, can be expected to produces the largest difference 
with respect to the viscous DNS. 

Two other popular 2D turbulent model were also tested along the same
line. In the first one (HDNS), the viscous term 
$\nu\Delta\Omega$ of the Navier-Stokes equation is replaced
by a "hyper-viscous" term $\nu_p\Delta^p\Omega$. In our simulations,
we took $p=8$ and $\nu_p=10^{-18}$. The 
second model is the Anticipated Potential
Vorticity Model (APVM) developed by Sadourny and Basdevant \cite{sadourny85}.
Both models are very cheap, taking only about 2 minutes
of computational time. They are also less accurate, as can be
seen from both the Fig. \ref{fig:dtvortls}, and the table. In the HDNS $64^2$, the
correlation coefficient is only about 50 percent in the end of
the simulation. For the APVM, it is higher (about 75 percent), but still
lower than our "minimal model''. The energy spectra of the 4 simulations
can also be compared. This is done in Fig. \ref{fig:dtspect}.
 Our model develops an energy spectrum very close to the APVM one, and
slightly less steep than the DNS at scales close to the cut-off.
This is because at this scale, viscous effects start being felt 
and tend to bend the spectrum. The HDNS spectrum is much steeper near 
the cut-off than both
the reference DNS and the two other models.

\subsection{Forced turbulence}

\begin{figure}[hhh]
\centerline{\psfig{file=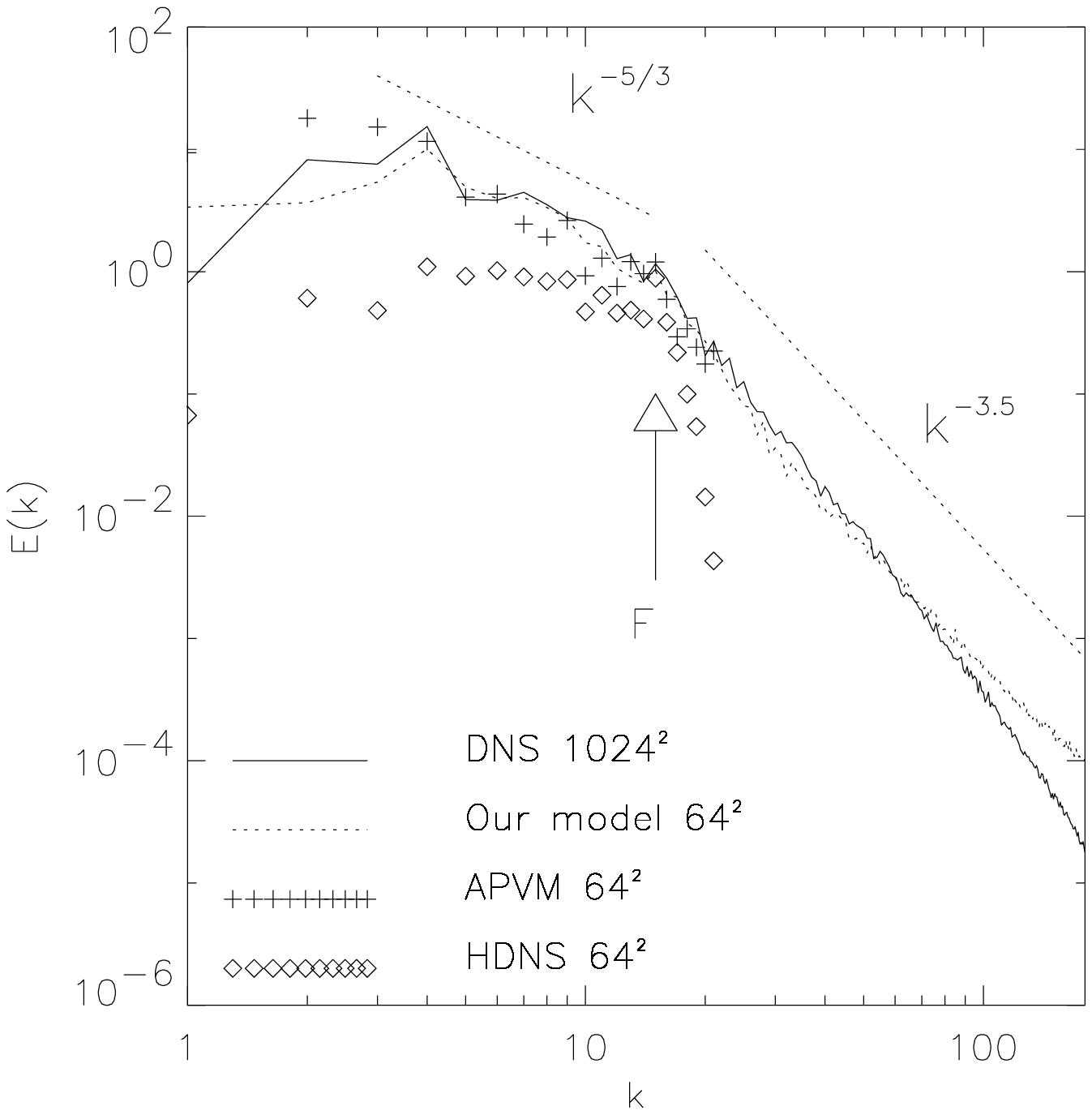,width=10cm}}
\caption[]{Comparison of the energy spectra of forced turbulence 
after 150 turnover times computed by different methods
(our model refers to the run M0 from the table 2)}
\label{fig:dtspectf}
\end{figure}

Similar simulations were performed 
in the case of forced turbulence. 
 In such a case, our model was ran with $M^2=64^2$ resolved-scale Fourier modes, 
and $N_p=512^2$ subgrid-scale Gabor modes, and no
additional viscosity. Run were also performed using a HDNS or 
the APVM model over $64^2$ Fourier modes. 
 The initial condition is a vorticity field with an energy spectrum
 concentrated at the forced wavenumber (k=15) and with a small amount of 
total energy. The simulation was forced by keeping constant the energy of
 the mode \kk=(15,0). In this situation, the vorticity field is progressively
built via the stochastic forcing, which is itself strongly dependent on
the exact structure of the vorticity field (since it must aim at keeping 
one mode constant). Due to the developing spectral cascades, the scale
interactions become less nonlocal
and  we may then expect 
sub-dominant terms to play an enhanced role (with respect to
the decaying case). Indeed, we have 
observed that the vorticity fields 
in the DNS and in the model do not exactly correspond: both simulations
display similar small scale intense vortices, but they are not located
at the same places. This effect can also be seen more clearly in the 
spectra. 
They are shown in fig. \ref{fig:dtspectf}.
%
%
%
%
Due to the chaoticity, they all differ at the mode $k=1$ which is 
the most sensitive to
 the exact position of each individual vortex. At smaller scales, marked 
differences appear between the models. Clearly, the HDNS 
gives the worst result, with a large deficit of energy over all scales.
This is because in this case, the forcing scale is very close to 
the scale at which the dissipation takes place, and most of the
energy is dissipated before the inverse cascade to larger scales can occur.
The two other models, which do not introduce an explicit dissipation at 
the cut-off scale, perform much better. The APVM model tends to overestimate
the rate of energy at the largest scale and to underestimate it
in the inertial range. Our model slightly underestimates the
amount of energy at the largest scale, but gives results very close
to the DNS in the inertial range of scale. At scales smaller than the cut-off,
both the DNS and our model produce an energy spectra steeper than the 
$k^{-3}$ law. At the largest wavenumber, the beginning of the viscous range
is clearly visible in the DNS but not in our model.
This is again an indication of the large Reynolds number achieved by our 
model. As for computational times, the performances are similar to the
decaying case. 
\clearpage
\section{Discussion}

We developed a new dynamic model of subgrid-scale 
turbulence based on a simple hypothesis
 about the subgrid-scale evolution. 
This approach is different from traditional  turbulent models
 since
 our model provides expressions of the turbulent Reynolds sub-grid stresses
 via estimates of the sub-grid velocities rather than velocities correlations.
The subgrid-scale dynamic is given by a linear equation, describing the
advection of subgrid-scale wave-packets with the mean flow. This feature allows
a reduction of the time step used in the simulation, via the use 
of a pseudo-Lagrangian method. We thereby achieved a reduction of
the computational time by  a factor $150$ in the typical cases
we considered. Our method can also be used for Large Eddy Simulation
strategies, in which only the large scales are computed. This allows
a large memory savings, and an additional reduction of the computational time 
(typically a factor 10 in the case we considered). The resulting 
simulation is more costly than a traditional LES simulation, based 
 on Fast Fourier Transform algorithms (like 
hyperviscosity or vorticity dissipative schemes). It could however become 
more competitive for more complicated geometries, where finite difference
schemes become more appropriate than FFT. This would make 
LES models based on hyperviscous schemes much more costly, while 
our method would proportionally keep at the same computational 
performance.

In the present paper, only 2D turbulence has been considered. As is well known,
2D turbulence is very special, because there is no vortex stretching. One
may therefore wonder whether our model could be applied to 3D flows. To
answer this question, Laval and Dubrulle \cite{laval00a} have recently 
performed a numerical analysis of the hypotheses pertaining our turbulent
model. The main hypothesis is that the subgrid dynamics can be very
well approximated by an equation in which non-local terms are 
retained, while local terms (non-linear in the subgrid velocities)
are replaced by a turbulent viscosity. The numerical analysis, 
performed via a numerical simulation of the coupled resolved and 
approximate subgrid equations indeed showed that such hypothesis is
valid, insofar as both the energy spectra, the structures and the 
statistical properties of such a model were very close to that of a DNS.
Even so this analysis can only be performed at a rather moderate 
Reynolds number (80 to 200, based on the Taylor scale), we find this
analysis encouraging.

\appendix


\section{Derivation of the subgrid scale equation in (\xx,\kk) space}
\label{sec:derivss}
Some properties of the GT will be useful in the sequel. They are 
(see~\cite{nazarenko99,dubrulle99b}
 for details):

\begin{eqnarray}
\widehat{\partial_i\uupr}(\xx,\kk,t) 
& = & \partial_i \hat{\uupr}(\xx,\kk,t) \\
                                     & = & i k_i \hat{\uupr}(\xx,\kk,t) +
 O({\epsilon}^*)
\label{eq:derivgab}
\end{eqnarray}

\begin{equation}
 \widehat{\UU\wpr}(\xx,\kk,t) \simeq  \UU(\xx,t) \hat{\wpr}(\xx,\kk,t) 
+ i \; \left(\grad_x\cdot \grad_k\right)\UU \wpr.
\end{equation}
Let us derive the GT of the equation:
\begin{equation}
\partial_t \wpr + div(U_j \wpr)  = F(\xx,t)+\nu_t \Delta \wpr.
\end{equation}
Using (\ref{eq:derivgab}), we find that the GT of the viscous term is 
$-\nu_t k^2 \hat \wpr$. 
The GT and the time derivative commute, so that
the GT of the first term of the lhs gives :
\begin{equation}
\widehat{\partial_t \wpr}  = \partial_t \;  \hat{\wpr}
\end{equation}
Using the space derivative property (\ref{eq:derivgab}), the GT of the second terms can be developed into:
\begin{eqnarray}
\widehat{\dr_j U_j \wpr} & = & \widehat{U_j \dr_j \wpr} \\
\nonumber & \simeq & U_j \; \widehat{\dr_j \wpr} + i \, \dr_l U_j \frac{\dr}{\dr k_l}
                                            \widehat{\dr_j \wpr} \\
\nonumber & \simeq & U_j \dr_j \hat{\wpr} + i \, \dr_l U_j \frac{\dr}{\dr k_l}
                                            \left( i k_j
                                            \hat{\wpr} \right) \\
\nonumber & \simeq & U_j \dr_j \hat{\wpr} + i \, \dr_l U_j k_j
                                            \frac{\dr}{\dr k_l}
                                            \hat{\wpr} - \hat{\wpr}\dr_j U_j                                   
\label{eq:tguwpr} 
\end{eqnarray} 
Using the incompressibility ($\dr_j U_j = 0$), we finally obtain :
\begin{equation}
 D_t \, \hat{\wpr}(\xx,\kk,t) = \hat{F}(\xx,t)-\nu_t k^2 \hat
\wpr (\xx,\kk,t)
\label{eq:consvss}
\end{equation}
with : \\
\begin{eqnarray}
 D_t       & = & \partial_t + \UU \cdot \grad -\grad_x\left(\UU\cdot\kk\right)
\cdot\grad_k, 
\label{eq:vortsstseg} \\
\end{eqnarray} 

\section{Reconstruction formulae for the correlations}
\label{sec:analreynolds}
Using the formula \ref{eq:derivgab} for the GT of space derivative, one can derive the subgrid scale velocity field
 in the (\xx,\kk) space with respect to the subgrid scale vorticity:
\begin{eqnarray}
\hat{\wpr}(x,y,p,q,t) & = & \widehat{\partial_x v^{\prime}}(x,y,p,q,t) - \widehat{\partial_y
u^{\prime}}(x,y,p,q,t) \\
\nonumber & = & \partial_x \hat{v^{\prime}}(x,y,p,q,t) - \partial_y \hat{u^{\prime}}(x,y,p,q,t) \\
\nonumber & = & i \, p \, \hat{v^{\prime}}(x,y,p,q,t) - i \, q \, \hat{u^{\prime}}(x,y,p,q,t) +
O({\epsilon}^{\star}) ,
\end{eqnarray}
where $\kk = (p,q)$ and $\xx = (x,y)$. Reversing the formula, we obtain the expression of the Gabor Transform of the velocity:

\begin{eqnarray}
\hat{u^{\prime}}(x,y,p,q,t) & = & \frac{iq}{p^2+q^2} \, \hat{\wpr}(x,y,p,q,t) +
O({\epsilon}^{\star}),
\label{eq:defvitug} \\
\hat{v^{\prime}}(x,y,p,q,t) & = & \frac{-ip}{p^2+q^2}\, \hat{\wpr}(x,y,p,q,t) +
O({\epsilon}^{\star}).
\label{eq:defvitvg}
\end{eqnarray}
Consider now the following expression:
\begin{eqnarray}
\nonumber &   & \int \frac{1}{2} \left[ \hat{\upr}(\xx, \kk,t) \hat{\wpr}(\xx,-\kk,t) + 
                    \hat{\upr}(\xx,-\kk,t) \hat{\wpr}(\xx, \kk,t) \right] d\kk \\
\nonumber & = & \int \Re \left[ \hat{\upr}(\xx, \kk,t) \hat{\wpr}(\xx,-\kk,t) \right] d\kk \\
\nonumber & = & \int \Re \left[ \int f({\epsilon}^{\star} (\xx-\xxpr)) e^{i\kk(\xx-\xxpr)}
\upr(\xxpr,t) d\xxpr \int f({\epsilon}^{\star} (\xx-\xxse)) e^{i\kk(\xx-\xxse)}
\wpr(\xxse,t) d\xxse \right] d\kk \\
          & = & \int f({\epsilon}^{\star} (\xx-\xxpr)) f({\epsilon}^{\star} (\xx-\xxse))
\upr(\xxpr,t) \wpr(\xxse,t) \left( \int e^{i\kk(\xxse-\xxpr)} d\kk \right) d\xxpr d\xxse .
\end{eqnarray}
Using the definition of the Dirac function,
\begin{equation}
 \frac{1}{(2\pi)^2} \int e^{i\kk(\xxse-\xxpr)} d\kk = \delta(\xxpr - \xxse),
\end{equation}
and the fact that $f^2=G$, one simply gets
\begin{equation}
\frac{1}{(2 \pi)^2} \int \Re \left[ \hat{\upr}(\xx, \kk,t) \hat{\wpr}(\xx,-\kk,t) \right] d\kk=
 \overline{\upr \wpr}(\xx,t) .
\label{eq:defmoyppp}
\end{equation}
We may proceed along the same line for 
the average of the non-linear product of a large scale field with a subgrid
 scale field. Using the definition of the average  $\overline{\UU \wpr}(\xx,t)$ can be written as follows,
\begin{equation}
 \overline{\UU \wpr}(\xx,t)  =  \int f^2({\epsilon}^{\star} (\xx-\xxpr)) \UU(\xxpr,t)\wpr(\xxpr,t) d\xxpr.
\label{eq:defmoyp2a}
\end{equation}
Using the Taylor development of $U$ with respect to $\xx$ at the first order,
 we obtain the first order approximation of this term:
\begin{eqnarray}
 \overline{\UU \wpr}(\xx,t) &  = & (\UU \overline{\wpr})(\xx) + \int
 f^2({\epsilon}^{\star} (\xx-\xxpr)) (\xx-\xxpr) \grad \UU(\xxpr,t)
 \wpr(\xxpr,t) d\xxpr \nonumber \\
 & = & (\UU \overline{\wpr})(\xx) + O({\epsilon}^{\star})
\label{eq:defmoyp2b} .
\end{eqnarray}
If we now apply the definition of the average for a product of two subgrid
 scales field
 (\ref{eq:defmoyppp}) with the quantities $1$ et $\wpr$, we finally obtain:
\begin{eqnarray}
 \overline{\UU \wpr}(\xx,t) &= &\UU \frac{1}{(2 \pi)^2} \int \Re
 \left[\hat{\wpr}(\xx,\kk,t) \hat{1}(\xx,-\kk,t) \right] d\kk + O({\epsilon}^{\star}) \\
&=& \UU \frac{1}{(2 \pi)^2} \int \Re \left[\hat{\wpr}(\xx,\kk,t) \hat{f}(k) \right] d\kk + O({\epsilon}^{\star}) .
\label{eq:defmoyp2c}
\end{eqnarray}

\section{Numerical computation of the Reynolds Stress components}
\label{sec:numreynolds}
The subgrid scale field in physical space can be obtained
 from their discrete formula in the (\xx,\kk) by an integration with respect to $\kk$:
\begin{eqnarray}
\wpr(\xx,t) & = & \frac{1}{(2\pi)^2\;f(0)} \int \hat{\wpr}(\xx,\kk,t) d\kk \\
\nonumber   & = & \frac{1}{(2\pi)^2\;f(0)} \int \sum_{\alpha=1}^{2N_p} \hat{\sigma}_{\alpha}(t)
\; S_{\xx}(\xx-\xx_{\alpha}(t)) \; \delta(\kk-\kk_{\alpha}(t)) \, d\kk \\
\nonumber    & = & \frac{1}{f(0)}  \sum_{\alpha=1}^{2N_p} \hat{\sigma}_{\alpha}(t)
\; S_{\xx}(\xx-\xx_{\alpha}(t)) \\
\nonumber    & = & \frac{1}{f(0)} \left\{  \sum_{\alpha_+=1}^{N_p} \hat{\sigma}_{\alpha_+}(t)
\; S_{\xx}(\xx-\xx_{\alpha_+}(t)) +  \sum_{\alpha_-=1}^{N_p} \hat{\sigma}_{\alpha_1}(t)
\; S_{\xx}(\xx-\xx_{\alpha_-}(t)) \right\} ,
\label{eq:rebuiltxp}
\end{eqnarray}
where
\begin{equation}
 \sum_{\alpha=1}^{2N_p} = \sum_{\alpha_+=1}^{N_p} + \sum_{\alpha_-=1}^{N_p} ,
\end{equation}
$\sum_{\alpha_+=1}^{N_p}$ means sum over half the particles with the wavenumber $\kk_\alpha$  and $\sum_{\alpha_-=1}^{N_p}$ is the sum over particles with an opposite wavenumber $-\kk_\alpha$. We choose the ``positive'' particles $\alpha_+$ with $\palp > 0$ and $\kk_\alpha = (\palp,\qalp)$. Because $\xx_{\alpha_+} = \xx_{\alpha_-}$, $\wpr(\xx,t)$ can be written:
\begin{equation}
\wpr(\xx,t) = \frac{1}{f(0)} \sum_{\alpha_+=1}^{N_p} \left
( \hat{\sigma}_{\alpha_+}(t) + \hat{\sigma}_{\alpha_-}(t) \right) S_{\xx}(\xx-\xx_{\alpha_+}(t)) .
\end{equation}
Since $\wpr(\xx,t)$ is real, the following property holds:
\begin{equation}
\hat{\sigma}_{\alpha_+}(t) = \hat{\sigma}^{*}_{\alpha_-}(t)
\label{eq:conjug}
\end{equation}
and the formula used to rebuild the vorticity field in physical space is :
\begin{equation}
\wpr(\xx,t) = \frac{2}{f(0)} \; \sum_{\alpha_+=1}^{N_p}
\Re \left[ \hat{\sigma}_{\alpha_+}(t) \right] \; S_{\xx}(\xx-\xx_{\alpha_+}(t)) \\
\label{eq:wprnum}
\end{equation}
where $\Re \left[ \hat{\sigma}_{\alpha_+}(t) \right]$ is the real part of 
$\hat{\sigma}_{\alpha_+}(t)$. The same developments can be
made for the two velocity component, using (\ref{eq:defvitug}) and (\ref{eq:defvitvg}):
\begin{eqnarray}
\upr(\xx,t) & = & \frac{2}{f(0)} \; \sum_{\alpha_+=1}^{N_p} \;
\frac{-\qalp}{\pqalp} \Im \left[\hat{\sigma}_{\alpha_+}(t)\right] \; S_{\xx}(\xx-\xx_{\alpha_+}(t)) \\
\vpr(\xx,t) & = & \frac{2}{f(0)} \; \sum_{\alpha_+=1}^{N_p} \;
\frac{\palp}{\pqalp} \Im \left[\hat{\sigma}_{\alpha_+}(t)\right] \; S_{\xx}(\xx-\xx_{\alpha_+}(t)) 
\end{eqnarray}
where $\Im \left[ \hat{\sigma}_{\alpha_+}(t) \right]$ is now the Imaginary part  of $\hat{\sigma}_{\alpha_+}(t)$.\\

The previous expressions can be used to compute velocity correlations.
Using the analytical definition of these terms with respect the the subgrid scales field in (\xx,\kk) space (\ref{eq:defmoyppp},\ref{eq:defmoyp2c}), and using the definition of the subgrid scale discretization (\ref{eq:rebuiltxk}), we
get:
\begin{eqnarray}
\overline{\upr \vpr}(\xx,t) & = & \frac{1}{(2 \pi)^2} \int \Re \left
[ \hat{\upr}(\xx, \kk,t) \hat{\vpr}(\xx,-\kk,t) \right] d\kk \label{eq:uvbp}\\
\nonumber & = & \frac{1}{(2 \pi)^2} \int \Re  \left[ \left
(  \sum_{\alpha=1}^{2N_p} \frac{i q}{p^2+q^2} \,
\hat{\sigma}_\alpha \, S_\xx(\xx-\xx_\alpha) \delta(\kk-\kk_\alpha) \right) \right. \\
\nonumber &  & \hspace{2.05cm} \left.
\left( \sum_{\beta=1}^{2N_p}\frac{+i p}{p^2+q^2} \, \hat{\sigma}_\beta \,
S_\xx(\xx-\xx_\beta) \delta(-\kk-\kk_\beta) \right) \right] d\kk  .
\label{eq:defuvnum}
\end{eqnarray}
If all the wave-packets have a different wavenumber (at least on a domain equal to the ``support'' of  $S_{x}$),  eq \ref{eq:defuvnum} can be written:
\begin{equation}
\overline{\upr \vpr}(\xx,t) = 2 \sum_{\alpha_+=1}^{N_p} \Re \left[ \frac{-
\qalp \palp  }{(\pqalp)^2} \hat{\sigma}_{\alpha_+} \hat{\sigma}_{\alpha_-}  S^2_\xx(\xx-\xx_{\alpha_+}) \right] .
\end{equation}
Using the fact that $\hat{\sigma}_{\alpha_-}$  and $\hat{\sigma}_{\alpha_+}$ are complex 
conjugate, we finally get:
\begin{eqnarray}
\nonumber \overline{\upr \vpr}(\xx,t) = 2 \sum_{\alpha_+=1}^{N_p} \frac{-
\qalp \palp }{(\pqalp)^2} |\hat{\sigma}_{\alpha_+}|^2
S^2_\xx(\xx-\xx_{\alpha_+}) ,\\
\overline{{\upr}^2}(\xx,t) = 2 \sum_{\alpha_+=1}^{N_p} \frac{+
\qalp^2 }{(\pqalp)^2} |\hat{\sigma}_{\alpha_+}|^2  S^2_\xx(\xx-\xx_{\alpha_+}) ,
\label{eq:rebussvss} \\
\nonumber \overline{{\vpr}^2}(\xx,t) = 2 \sum_{\alpha_+=1}^{N_p} \frac{+
\palp^2 }{(\pqalp)^2} |\hat{\sigma}_{\alpha_+}|^2  S^2_\xx(\xx-\xx_{\alpha_+}) .
\end{eqnarray}
The non-linear term involving the resolved
 scale velocity field can also be written in terms of wave-packets
 coordinates. Using eq. (\ref{eq:defmoyp2c}) and the definition of discretization (\ref{eq:rebuiltxk}) , one can write:
\begin{eqnarray}
\overline{\UU \wpr}(\xx,t) & \simeq &  \frac{\UU}{(2 \pi)^2} \int \Re
 \left[\hat{\wpr}(\xx,\kk,t) \hat{1}(\xx,-\kk,t) \right] d\kk \nonumber\\
& \simeq &  \frac{\UU}{(2 \pi)^2} \int \Re
 \left[\hat{\wpr}(\xx,\kk,t) f(-\kk) \right] d\kk \nonumber \\
& \simeq & \frac{\UU}{(2 \pi)^2} \int  \Re
 \left[ \sum_{\alpha=1}^{2 N_p} \hat{\sigma_\alpha} f^*(\kk)  S_\xx(\xx-\xx_\alpha)
\delta(\kk-\kk_\alpha)\right]  d\kk \nonumber \\
& \simeq &  2 \UU  \sum_{\alpha_+=1}^{N_p} \Re
 \left[\hat{\sigma_{\alpha_+}} f^*(\kk_{\alpha_+}) \right]
 S_\xx(\xx-\xx_{\alpha_+}) .
\label{eq:rebulswss}
\end{eqnarray}

\section{Choice of the filter}
\label{sec:filter}

The choice of the
 interpolating function $S_x(\xx)$  (eq. \ref{eq:picfuncx})
dictates the shape of the filter and, therefore, of
 the unknown function $f(x)$ in the definition of the Gabor Transform. 
The proof proceeds via the quantity $\overline{{\wpr}^2}(\xx,t)$, 
by equating 
the formula of the filter discretized on a regular grid with a cell size $\Delta \xx$ and the formula \ref{eq:wprnum}:
\begin{eqnarray}
\overline{{\wpr}^2}(\xx,t) & = & \sum_{i} {\wpr}^2(\xx_i) \, f^2(\xx-\xx_i) \, (\Delta
\xx)^2 \\
& = & 2 \sum_{\alpha_+=1}^{N_p} |\sigma_{\alpha_+}|^2 \, S^2_\xx(\xx-\xx_{\alpha_+})
\end{eqnarray}
For this equality to be valid at all
grid point $X_i$, the following link between $f$ and $S$ must hold:
\begin{equation}
f^2(\xx) = C \times S^2_\xx(\xx).
\label{eq:deffilcond}
\end{equation}
Using now the normalization:
\begin{eqnarray}
S_\xx(0) &=& 1. \\
\int f^2(\xxpr) d \xxpr &=& 1.
\end{eqnarray}
the eq.  \ref{eq:deffilcond} becomes:
\begin{equation}
\D f (\xx)  =  f(0) \, S_\xx(\xx) \\
\label{eq:deff}
\end{equation}
with
\begin{equation}
\D f(0) =   \frac{1}{\sqrt{\D \int S^2_\xx(\xxpr) d\xxpr}} .
\label{eq:deff0}
\end{equation}
%


\bibliographystyle{mybib}

\bibliography{../../Biblio/biblio}

\end{document}